\documentclass[conference]{IEEEtran}
\usepackage{graphicx}
\usepackage{booktabs}
\usepackage{multirow}
\usepackage{cite}
\usepackage{url}
\usepackage{array}
\usepackage{tabularx}
\usepackage{microtype}
\usepackage{amsmath}
\usepackage{algorithm}
\usepackage{algpseudocode}
\usepackage{xcolor}
\usepackage{listings}

\definecolor{kwblue}{RGB}{0,0,180}
\definecolor{strgreen}{RGB}{0,128,0}
\definecolor{cmtgray}{RGB}{120,120,120}
\definecolor{mutred}{RGB}{200,0,0}

\lstdefinestyle{paperpy}{%
  language=Python,
  basicstyle=\ttfamily\scriptsize,
  keywordstyle=\color{kwblue},
  stringstyle=\color{strgreen},
  commentstyle=\color{cmtgray}\itshape,
  showstringspaces=false,
  breaklines=true,
  columns=fullflexible,
  keepspaces=true,
  aboveskip=2pt,
  belowskip=2pt,
  xleftmargin=2pt,
  framerule=0pt,
  moredelim=[is][\color{mutred}]{<m>}{</m>},
  moredelim=[is][\bfseries]{<b>}{</b>},
}
\lstdefinestyle{paperpycompact}{%
  style=paperpy,
  aboveskip=0pt,
  belowskip=0pt,
  xleftmargin=0pt,
}

\usepackage[hidelinks]{hyperref}
\begin{document}

\title{Deep4ge: DNN Training Trajectories for Fault Detection and Diagnosis}
\author{\IEEEauthorblockN{Sigma Jahan}
\IEEEauthorblockA{Dalhousie University, Canada\\
\texttt{sigma.jahan@dal.ca}}
}

\maketitle

\begin{abstract}
Deep learning systems often fail due to subtle implementation faults that alter training behavior. Recent work has studied how to detect and diagnose such failures from changes observed across training epochs. However, the software engineering community still lacks a public dataset of per-epoch training runs with documented fault history, feature extraction details, and clear reuse support for fault detection and diagnosis tasks. We present Deep4ge, a controlled benchmark of 14,227 training runs generated from 59 adapted TensorFlow/Keras deep neural network (DNN) programs collected from Stack Overflow. We generated faulty variants using 27 source-code transformations that introduce known faults across seven categories. The dataset contains 9,845 faulty runs and 4,382 correct baseline runs. For each run, we record 4 evaluation metrics and 26 features that measure training behavior at every epoch. These features capture weights, gradients, activations, accuracy and loss trends, learning rate, and hardware use. Deep4ge supports binary fault detection, multiclass fault diagnosis, and early fault prediction from partial training runs. We release the dataset and fault-injection framework at \url{https://doi.org/10.5281/zenodo.20337241}.
\end{abstract}


\section{Introduction}

Deep neural network (DNN) components are now built into software systems, making training-time failures a major concern. Unlike traditional faults that produce explicit errors, DNN faults often appear as silent symptoms during training. These symptoms include unstable optimization, vanishing or exploding gradients, and poor generalization. Debugging such faults therefore requires observing how training evolves over time rather than examining only the source code or final output. This methodological shift warrants datasets that record training behavior throughout the learning process.

Recent work has proposed approaches for debugging DNN systems, including AutoTrainer~\cite{Zhang2021AutoTrainer}, UMLAUT~\cite{Schoop2021Umlaut}, DeepLocalize~\cite{Wardat2021DeepLocalize}, DeepDiagnosis~\cite{Wardat2022DeepDiagnosis}, and DeepFD~\cite{Cao2022DeepFD}. These approaches are typically evaluated on a limited number of programs or on study-specific injected faults, and the corresponding training runs are often not released in a form that supports reuse or systematic re-evaluation. Mutation frameworks such as DeepCrime~\cite{Humbatova2021DeepCrime} apply source-code transformations derived from real faults, but they mainly support mutation analysis and operator evaluation. As a result, support for studying training-time faults in deep learning systems remains limited.

In this work, we address this gap with Deep4ge. We collect TensorFlow/Keras programs from Stack Overflow, adapt them to run deterministically, apply these transformations, and log measurements at every epoch to form training trajectories. Deep4ge contains 14,227 training runs across seven fault categories and three architecture families. As a controlled benchmark, it can be extended by adding new transformations, programs, or features and regenerating the associated metadata. In summary, we make the following contributions:
\begin{itemize}
\item A labeled dataset of 14,227 per-epoch DNN training runs across seven fault categories.
\item An open-source fault-injection framework that applies these transformations and logs the resulting training runs with validation scripts.
\item Proof-of-concept fault diagnosis baselines on the dataset.
\end{itemize}

\section{Dataset Construction}
\subsection{Program Selection}

\begin{figure}[htbp]
\centering
\includegraphics[width=\linewidth]{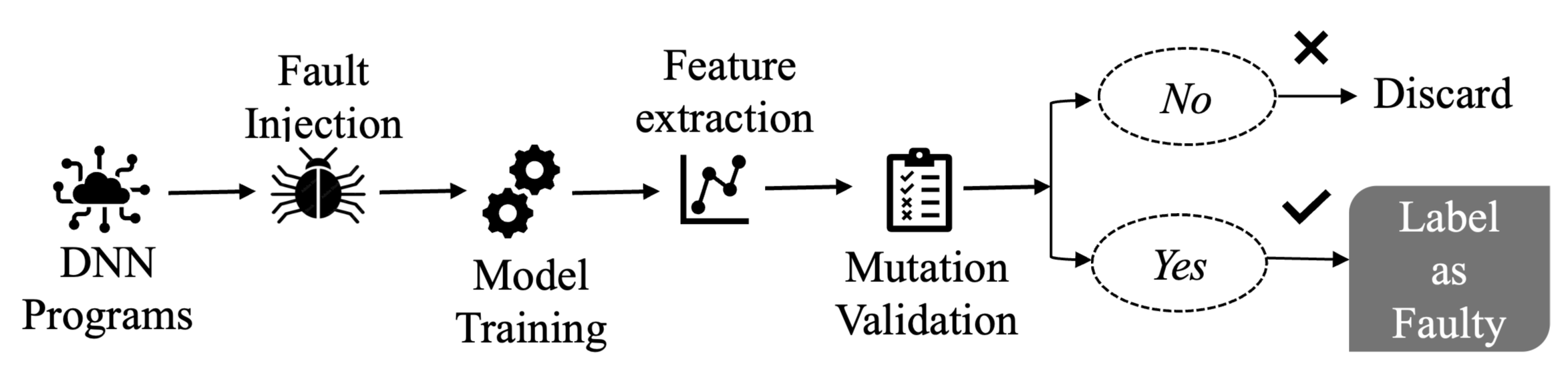}
\caption{Workflow of Deep4ge construction}
\label{fig:pipeline}
\end{figure}

Deep4ge is based on developer-posted code. We collected 59 TensorFlow/Keras DNN programs from Stack Overflow (SO) that include the training code. We applied three criteria to select programs. First, the post must contain a complete model definition with data loading, compilation, and training. Second, the program must be repairable to run deterministically. Third, the architecture must be in scope. We then manually fixed each program to ensure it runs successfully and instrumented it to record per-epoch features during training. To reduce non-deterministic variability across runs, we fixed random seeds at the framework and operator level and ensured consistent dataset splits and initialization settings during execution. The seeding setup allows the dataset to be regenerated under the same configurations, supporting comparisons across fault categories.

\textit{Collection date and identifiers.} The artifact records the SO question identifier for each program and the collection date. Each program corresponds to a distinct SO question ID.

\textit{Architecture coverage.} The dataset includes feedforward, convolutional, and recurrent neural networks.

\textit{Licensing and attribution.} The artifact stores program metadata and links each program to its originating SO question, with license and attribution information included in the release.

\subsection{Fault Injection}
Deep4ge focuses on faults in the model and training code. Starting from each corrected program, we generate faulty variants by applying one of 27 source-level mutation operators to each variant. The operators are grouped into seven fault categories. The operator set combines a subset of DeepCrime operators with additional layer-level operators introduced in Deep4ge. From DeepCrime, we retained operators covering training configuration, activation, loss, optimizer, weight initialization and bias, and regularization. We excluded operators that modify training data or process-level control, as these are outside our scope. The Deep4ge framework adds the full set of layer-related mutation operators not supported in prior work~\cite{Humbatova2021DeepCrime}.

The framework implements 27 mutation operators covering seven model and training fault categories represented in the literature~\cite{Humbatova2021DeepCrime, islam_comprehensive_2019}. Each released run is generated using one operator and receives that operator's category in \texttt{manifest.csv}. A seed program can contribute separate variants from multiple categories, but each run has one category label. The diagnosis benchmark is therefore a seven-class, single-label task and does not evaluate multilabel diagnosis.

\subsection{Training and Feature Extraction}
We train each mutated program and its correct baseline. Each run logs, per epoch, an epoch index, 4 metrics (training and validation loss and accuracy), and 26 dynamic features. These comprise four weight measures (large-weight count, unchanged-mean and unchanged-standard-deviation flags, and NaN count), four accuracy/loss trends (gap, oscillation, decrease, and increase), four activation measures (dying ReLU, saturation, mean, and standard deviation), and ten gradient measures (vanishing/exploding flags, NaN count, mean, standard deviation, minimum, maximum, and median layer norms, and per-weight mean and standard deviation). The remaining four record learning rate, CPU use (\%), peak GPU memory (MB), and system-memory use (\%). GPU memory is zero when no GPU is available. We apply each operator by rewriting the program's abstract syntax tree, which also inserts the logging callback (see Fig.~\ref{fig:codepair}).

\begin{figure*}[htbp]
\centering
\begin{minipage}[t]{0.495\textwidth}
\centering
\text{(a) Seed training fragment}
\lstset{style=paperpycompact}
\begin{lstlisting}
<b>from CustomCallback import EnhancedLoggingCallback</b>
model = Sequential()
model.add(Dense(4, input_dim=2,
    kernel_initializer="glorot_uniform"))
model.add(Activation("sigmoid"))
model.add(Dense(1))
model.add(Activation("sigmoid"))
model.compile(loss="mse",
    optimizer=Adam(learning_rate=1e-3), metrics=["accuracy"])
<b>cb = EnhancedLoggingCallback(train_dataset, callback_filename)</b>
model.fit(x_train, y_train, epochs=50, batch_size=16,
    validation_data=(x_test, y_test), <b>callbacks=[cb]</b>)
\end{lstlisting}
\end{minipage}\hfill%
\begin{minipage}[t]{0.495\textwidth}
\centering
\text{(b) AST operator application}
\lstset{style=paperpycompact}
\begin{lstlisting}
tree = ast.parse(source)
tree = ModifySavePath(suffix).visit(tree)
<b>tree = ModifyCallbackFilename(csv_path).visit(tree)</b>
<b>OpClass = OPERATOR_CLASSES[op_code]</b>
<b>tree = OpClass().visit(tree)</b>
ast.fix_missing_locations(tree)
code = compile(tree, "<ast>", "exec")
exec(code, {"__name__": "__main__"})
# example operator class
class ChangeLearningRate(MutationOperator):
    def visit_Call(self, call):
        self.generic_visit(call)
        if is_compile_call(call): <m>rewrite learning_rate</m>
        return call
\end{lstlisting}
\end{minipage}
\caption{AST-based mutation and feature logging (per-epoch)}
\label{fig:codepair}
\end{figure*}

\subsection{Mutation Validation}
We follow the killing criterion of DeepCrime~\cite{Humbatova2021DeepCrime}, which builds on the statistical mutation-killing notion of Jahangirova and Tonella~\cite{Jahangirova2020Mutation} and is also adopted in DEFault~\cite{Jahan2025DEFault}. Let $A_N$ and $A_M$ be the accuracy distributions of the original network $N$ and a mutant $M$ over $n$ retrainings on the test set $T$. The mutant is killed by $T$ when the difference between the two distributions is statistically significant with at least a medium effect:
\begin{equation}
\label{eq:iskilled}
\textsc{isKilled}(N, M, T) \iff d \geq \beta \;\wedge\; p < \alpha,
\end{equation}
where $d$ and $p$ are the effect size and significance value computed from $A_N$ and $A_M$.
We use Cohen's $d$ ($\beta = 0.5$), a Generalized Linear Model for the $p$-value ($\alpha = 0.05$), and $n = 15$. A mutant that the criterion does not kill is likely equivalent to the original program, so we drop it and do not include its runs in the release. For an operator $O$ with configuration space $C$, let $K(O, T) \subseteq C$ be the configurations of $O$ killed by $T$, and let $\mathrm{TrainS}$ denote the training set used to select configurations. The mutation score is
\begin{equation}
\label{eq:ms}
\mathit{MS}(O, T) = \frac{\lvert K(O, T) \cap K(O, \text{TrainS}) \rvert}{\lvert K(O, \text{TrainS}) \rvert}.
\end{equation}
We retained a mutation operator only if it was killable on at least one program of the same architecture. The operator set achieved mutation scores of 0.85 (FNN), 0.87 (CNN), and 0.83 (RNN). Finally, we normalize filenames, convert logged boolean values, and discard runs that did not complete a single epoch. A validation script checks consistency across all runs. Our replication package includes the construction scripts, allowing practitioners to reapply operators, retrain models, and regenerate the dataset from scratch.

\begin{algorithm}[htbp]
\caption{Deep4ge construction loop}
\label{alg:construction}
\footnotesize
\begin{algorithmic}[1]
\Require Programs $\mathcal{P}$, operators $\mathcal{O}$, retrainings $n$, seeds $\mathcal{S}$
\State $\mathcal{D} \gets \emptyset$
\Comment{collected runs}
\For{each program $P \in \mathcal{P}$}
  \State train $P$ for $n$ retrainings, log per-epoch values
  \State append the $n$ correct baseline runs to $\mathcal{D}$
  \For{each operator $O \in \mathcal{O}$}
    \State $M \gets$ apply $O$ to $P$
    \State train $M$ for $n$ retrainings with seeds $\mathcal{S}$, log per-epoch values
    \If{$\textsc{isKilled}(P, M, T)$}
      \Comment{Eq.~\ref{eq:iskilled}}
      \State append the $n$ faulty runs to $\mathcal{D}$, label by category of $O$
    \EndIf
  \EndFor
\EndFor
\State standardize, clean, and validate $\mathcal{D}$
\State build the metadata indexing every run in $\mathcal{D}$
\State \Return $\mathcal{D}$ and its metadata
\end{algorithmic}
\end{algorithm}

\section{Dataset Structure and Access}

\textbf{Dataset organization.} The dataset contains three top-level items: metadata, 59 adapted Stack Overflow programs, and training logs of faulty and correct runs.

\textbf{Schema.} Each entry in \texttt{manifest.csv} identifies the subset, SO ID, operator-derived fault category, run number, epoch count, and file path. Each run records, per epoch, an epoch index, four evaluation metrics, and 26 dynamic features. The data dictionary is included in the artifact.

\begin{table}[htbp]
\centering
\begin{minipage}[t]{0.48\columnwidth}
\centering
\caption{Dataset summary}
\label{tab:summary}
\resizebox{\linewidth}{!}{%
\begin{tabular}{@{}lr@{}}\toprule
Property & Value \\ \midrule
Total training runs & 14,227 \\
Faulty runs & 9,845 \\
Correct baseline runs & 4,382 \\
Unique Stack Overflow IDs & 59 \\
Total epoch records & 719,560 \\
Median epochs / run & 50 \\
\bottomrule\end{tabular}}
\end{minipage}\hfill%
\begin{minipage}[t]{0.50\columnwidth}
\centering
\caption{Fault counts by architecture}
\label{tab:arch}
\resizebox{\linewidth}{!}{%
\begin{tabular}{@{}lrrrr@{}}\toprule
Arch. & \#programs & \#runs & \#faulty & \#correct \\ \midrule
CNN & 13 & 2,683 & 2,037 & 646 \\
FNN & 16 & 3,792 & 2,555 & 1,237 \\
RNN & 30 & 7,752 & 5,253 & 2,499 \\
\bottomrule\end{tabular}}
\end{minipage}
\end{table}

\begin{table}[htbp]
\centering
\caption{Schema sample showing metadata \& per-run fields}
\label{tab:schema}
\scriptsize
\setlength{\tabcolsep}{4pt}
\begin{tabular}{@{}l l l@{}}
\toprule
\textbf{Field} & \textbf{Type} & \textbf{Description} \\
\midrule
\multicolumn{3}{c}{\textit{Metadata}}\\
\path{filename} & string & Per-run record filename \\
\path{subset} & enum & \path{buggy} or \path{correct} \\
\path{so_id} & string & Stack Overflow question ID \\
\path{fault_category} & enum & Operator category, or \path{none} \\
\path{is_faulty} & bool & True if the run is faulty \\
\path{run_number} & int & Index within a program--operator pair \\
\path{num_epochs} & int & Epochs recorded in the run \\
\path{csv_path} & path & Path to the per-run record \\
\midrule
\multicolumn{3}{c}{\textit{Per-run record (epoch-wise)}}\\
\path{epoch} & int & Epoch index \\
\path{train_loss}, \path{val_loss} & float & Training and validation loss \\
\path{train_acc}, \path{val_acc} & float & Training and validation accuracy \\
Feature fields & num/bool & 26 dynamic features (see data dictionary) \\
\bottomrule
\end{tabular}
\normalsize
\end{table}

\textbf{Availability.} The dataset is released under CC BY 4.0 and the mutation framework under MIT, archived on Zenodo with a citable DOI. Each run traces to its originating SO post and operator through the metadata. The release is version \texttt{v1.0.0} (commit \texttt{3f30d85e4f81}), with the repository at \url{https://github.com/SigmaJahan/deep4ge} and the archived release at \url{https://doi.org/10.5281/zenodo.20337241}.

\section{Dataset Characterization}

Deep4ge contains 14,227 training runs (9,845 faulty and 4,382 correct baselines) spanning 719,560 epoch records (median 50 epochs per run). The dataset covers seven fault categories and three architectures (see Table~\ref{tab:operator_taxonomy}).

\begin{figure}[htbp]
\centering
\includegraphics[width=\linewidth]{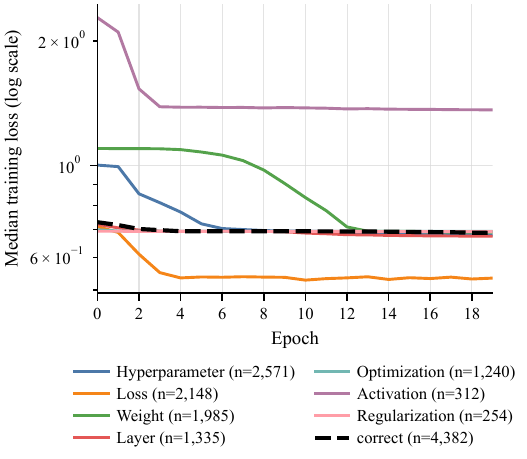}
\caption{Median per-epoch training loss by fault category on a log scale (first 18 epochs)}
\label{fig:loss_curves}
\end{figure}

\textbf{Training dynamics.} Within Deep4ge, the median training-loss trajectories form three broad behavioral groups (Fig.~\ref{fig:loss_curves}). Some categories clearly diverge from the correct baseline. The Activation trajectory does not converge within the first 18 epochs, and the Loss trajectory settles on a different numerical scale. Others start high but converge within the displayed range. Hyperparameter faults recover within about seven epochs, while Weight-initialization faults take until around epoch 13. Layer, Optimization, and Regularization faults closely track the baseline and are the hardest to distinguish from loss alone. These groupings show that the injected fault categories represented in Deep4ge differ in both when and how their median loss deviates, which the final epoch alone cannot capture. Layer mutations are concentrated in CNN and RNN programs, where the applicable operators target structural choices such as kernel size and LSTM layer count (Table~\ref{tab:cat_arch}).

\begin{table}[htbp]
\centering
\caption{Mutation operator taxonomy by fault category}
\label{tab:operator_taxonomy}
\scriptsize
\setlength{\tabcolsep}{4pt}
\begin{tabular}{@{}l l r l@{}}
\toprule
\textbf{Category} & \textbf{Operators} & \textbf{Logs} & \textbf{Exemplar fault} \\
\midrule
Hyperparameter & HBS, HLR, HNE, HDB & 2{,}571 & Bad learning rate \\
Loss           & FLC                & 2{,}148 & Wrong loss function \\
Weight         & WCI, WAB, WRB      & 1{,}985 & Bad weight init \\
Layer          & LKS, LCF, \ldots, LCO ($10$) & 1{,}335 & Wrong kernel size \\
Optimization   & OCH, OCG           & 1{,}240 & Optimizer swapped \\
Activation     & ACH, ARM, AAL      &    312  & ReLU removed \\
Regularization & RCD, RAW, RCW, RRW &    254  & Dropout set to 1.0 \\
\midrule
\textit{Total} & \textit{27 operators} & \textit{9{,}845} & \textit{+4{,}382 correct} \\
\bottomrule
\end{tabular}
\normalsize
\end{table}

\section{Detection and Diagnosis Benchmarks}
\label{sec:utility}

Deep4ge enables a wide range of studies on training-time faults. As a proof-of-concept, we report baselines on two tasks: \emph{fault detection}, which classifies runs as faulty or correct, and \emph{fault diagnosis}, which assigns one of seven fault categories to faulty runs. These baselines serve as a starting point and illustrate the analyses supported by the dataset.

\textbf{Experimental setup.} We compare two feature representations. The \emph{final-epoch} representation uses the 30 values recorded at the last epoch. The \emph{trajectory} representation summarizes each of the 30 features over the full run using six statistics (mean, standard deviation, minimum, maximum, last value, and linear slope), yielding 180 features. We evaluate Logistic Regression (LR) and Random Forest (RF)~\cite{Breiman2001RandomForests} using 5-fold group-aware cross-validation, grouping runs by SO program identifier so that runs from the same program do not appear in both training and test sets. This prevents leakage across runs from the same program and measures transfer to held-out Stack Overflow programs. We report F1, Macro-F1, balanced accuracy, and the Matthews correlation coefficient (MCC)~\cite{Matthews1975MCC, Chicco2020MCC}, along with 95\% bootstrap confidence intervals computed over 1{,}000 resamples. We consider always-faulty and stratified-random classifiers as trivial baselines.

\textbf{Final-epoch features.} The final-epoch features provide limited separation between faulty and correct runs. With final-epoch features, Random Forest achieves an F1 score of 0.816 on detection (Table~\ref{tab:representations}). Most runs in the dataset are faulty (9{,}845 of 14{,}227), so a classifier that labels all runs as faulty achieves an F1 score of 0.818. Random Forest performs comparably to the trivial baseline in terms of F1. To reduce sensitivity to class imbalance, we consider MCC and balanced accuracy, which equal 0 and 0.5 respectively for the trivial classifier. On these metrics, Random Forest improves modestly over the trivial baseline (MCC 0.150, balanced accuracy 0.540), while Logistic Regression achieves a higher MCC (0.184).

\textbf{All-epoch features.} Trajectory features provide a stronger basis than final-epoch representations. Using the full training trajectory instead of only the last epoch increases Random Forest detection MCC from 0.150 to 0.227, with non-overlapping confidence intervals ([0.132, 0.169] versus [0.209, 0.245]). Detection performance also improves in AUROC from 0.649 to 0.716 and in AUPRC from 0.804 to 0.842. The same pattern appears in diagnosis results, where Random Forest Macro-F1 increases from 0.355 to 0.381 and MCC increases from 0.346 to 0.401. Both models use per-feature statistics with no hyperparameter tuning. Under this evaluation, Deep4ge's per-epoch trajectories support higher detection and diagnosis performance than final-epoch information alone.

\begin{table}[t]
\centering
\caption{Detection \& diagnosis: final-epoch vs.\ trajectory features}
\label{tab:representations}
\resizebox{\columnwidth}{!}{%
\begin{tabular}{@{}lllrrrr@{}}\toprule
Task & Features & Model & F1 & Macro-F1 & Bal.Acc & MCC \\ \midrule
Detection & final-epoch & Always-faulty & 0.818 & 0.409 & 0.500 & 0.000 \\
Detection & final-epoch & Stratified & 0.693 & 0.497 & 0.497 & -0.006 \\
Detection & final-epoch & LogReg & 0.611 & 0.553 & 0.599 & 0.184 \\
Detection & final-epoch & RandForest & 0.816 & 0.508 & 0.540 & 0.150 \\
\midrule
Diagnosis & final-epoch & Stratified & 0.141 & 0.141 & 0.141 & -0.003 \\
Diagnosis & final-epoch & LogReg & 0.214 & 0.214 & 0.258 & 0.165 \\
Diagnosis & final-epoch & RandForest & 0.355 & 0.355 & 0.422 & 0.346 \\
\midrule
Detection & trajectory & Always-faulty & 0.818 & 0.409 & 0.500 & 0.000 \\
Detection & trajectory & Stratified & 0.693 & 0.497 & 0.497 & -0.006 \\
Detection & trajectory & LogReg & 0.693 & 0.583 & 0.599 & 0.185 \\
Detection & trajectory & RandForest & 0.825 & 0.553 & 0.568 & 0.227 \\
\midrule
Diagnosis & trajectory & Stratified & 0.141 & 0.141 & 0.141 & -0.003 \\
Diagnosis & trajectory & LogReg & 0.257 & 0.257 & 0.320 & 0.207 \\
Diagnosis & trajectory & RandForest & 0.381 & 0.381 & 0.465 & 0.401 \\
\bottomrule\end{tabular}}
\end{table}

\begin{figure}[htbp]
\centering
\includegraphics[width=0.96\linewidth]{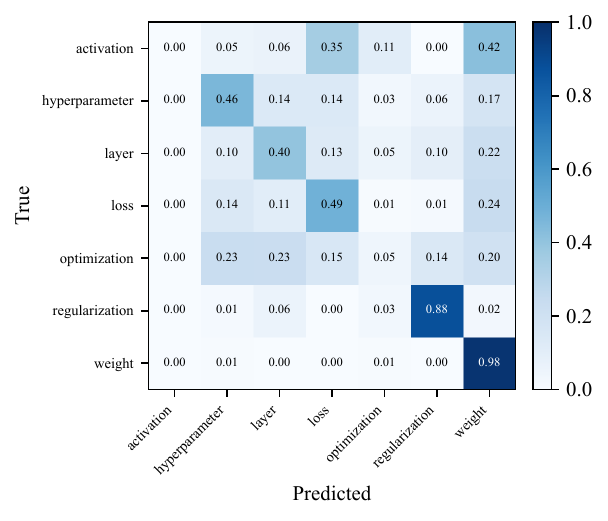}
\caption{Fault-diagnosis confusion matrix (normalized)}
\label{fig:confusion}
\end{figure}

\textbf{Per-category diagnosis.} With per-epoch feature trajectories, we observe variation in diagnosis performance across fault categories (Table~\ref{tab:per_class}). Weight-related faults achieve the highest performance, followed by Hyperparameter and Loss. In contrast, Optimization and Activation faults are the most difficult to diagnose. Fig.~\ref{fig:confusion} shows that Activation faults are frequently misclassified as Optimization, while Optimization faults are often misclassified as Layer. These patterns suggest that some fault categories may exhibit similar training dynamics. Deep4ge enables further analysis of such patterns and supports future work on more discriminative features for difficult fault categories.

\begin{table}[htbp]
\centering
\begin{minipage}[t]{0.46\columnwidth}
\centering
\caption{Fault diagnosis performance}
\label{tab:per_class}
\resizebox{\linewidth}{!}{%
\begin{tabular}{@{}lr@{}}\toprule
Fault Category & Per-class F1 \\ \midrule
activation & 0.000 \\
hyperparameter & 0.523 \\
layer & 0.384 \\
loss & 0.520 \\
optimization & 0.075 \\
regularization & 0.462 \\
weight & 0.700 \\
\bottomrule\end{tabular}}
\end{minipage}\hfill%
\begin{minipage}[t]{0.50\columnwidth}
\centering
\caption{Cross-architecture detection}
\label{tab:crossarch}
\resizebox{\linewidth}{!}{%
\begin{tabular}{@{}lrrr@{}}\toprule
Setting & F1 & Bal.Acc & MCC \\ \midrule
FNN & 0.823 & 0.563 & 0.210 \\
CNN & 0.827 & 0.566 & 0.218 \\
RNN & 0.820 & 0.559 & 0.198 \\
in-distribution & 0.825 & 0.568 & 0.227 \\
\bottomrule\end{tabular}}

\vspace{6pt}
\caption{Faulty runs by architecture}
\label{tab:cat_arch}
{\scriptsize
\setlength{\tabcolsep}{4pt}
\begin{tabular}{@{}lrrr@{}}
\toprule
\textbf{Category} & \textbf{FNN} & \textbf{CNN} & \textbf{RNN} \\
\midrule
Hyperparameter & 684 & 304 & 1{,}583 \\
Loss           & 586 & 516 & 1{,}046 \\
Weight         & 434 & 484 & 1{,}067 \\
Layer          & 460 & 282 &   593 \\
Optimization   & 270 & 360 &   610 \\
Activation     &  24 &   0 &   288 \\
Regularization &  97 &  91 &    66 \\
\bottomrule
\end{tabular}}
\end{minipage}
\end{table}

\textbf{Cross-architecture transfer.} We train RF detection models on two architecture families and test on the held-out third (Table~\ref{tab:crossarch}). The in-distribution model achieves an MCC of 0.227. Cross-architecture models achieve MCCs of 0.210 (FNN), 0.218 (CNN), and 0.198 (RNN), suggesting that performance loss from held-out architectures is small but nonzero and is slightly larger for RNN. This larger gap is consistent with differences between gradient flow across time steps in recurrent models and layer-wise propagation in FNN and CNN architectures~\cite{grossberg2013recurrent}, although our evaluation does not isolate this mechanism. Overall, some fault-detection features transfer across the three evaluated architecture families, while others remain architecture-specific.

\begin{figure}[htbp]
\centering
\includegraphics[width=\linewidth]{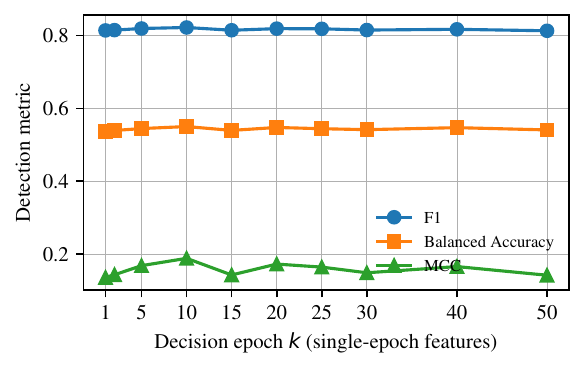}
\caption{Single-epoch fault detection by decision epoch (RF, group-aware CV)}
\label{fig:early_detection}
\end{figure}

\textbf{Early detection.} We evaluate whether fault detection is feasible from features extracted at a single training epoch. We repeat RF detection using only the feature values from epoch $k$, where $k \in \{1, 2, 5, 10, 15, 20, 25, 30, 40, 50\}$. MCC ranges from 0.14 to 0.19 across all values of $k$, compared to 0.227 for the in-distribution model that uses the full trajectory (Fig.~\ref{fig:early_detection}). No single epoch outperforms the others, and all single-epoch models underperform the full-trajectory model. Under this evaluation, Deep4ge's per-epoch trajectories support higher detection performance than single-epoch information alone.

\section{Related Work}

Deep4ge complements DEFault, our prior work on hierarchical and explainable DNN fault detection and diagnosis~\cite{Jahan2025DEFault}. DEFault used the dataset to evaluate detection and diagnosis techniques. This paper documents its construction and characterizes how per-epoch trajectories vary across fault categories and architectures. To the best of our knowledge, Deep4ge is the first public dataset that links adapted DNN programs with mutation-derived categories and corresponding per-epoch trajectories.

Mutation-testing work, including DeepMutation~\cite{Ma2018DeepMutation}, DeepMutation++~\cite{Hu2019DeepMutationPP}, and DeepCrime~\cite{Humbatova2021DeepCrime}, primarily focuses on operators and mutation analysis. Debugging techniques such as AutoTrainer~\cite{Zhang2021AutoTrainer}, UMLAUT~\cite{Schoop2021Umlaut}, DeepLocalize~\cite{Wardat2021DeepLocalize}, DeepDiagnosis~\cite{Wardat2022DeepDiagnosis}, and DeepFD~\cite{Cao2022DeepFD} evaluate detection strategies on less reusable datasets. Deep4ge instead links source programs, mutation-derived categories, and per-epoch features in one reusable benchmark.

\section{Use Cases and Limitations}
\textit{Use cases:} Deep4ge supports early fault detection from partial runs, comparison of detection and diagnosis techniques across categories and architectures, feature-importance analysis, and cross-architecture studies. Its trajectories, labels, and source programs support replication and extension of existing techniques.

\textit{Limitations:} Deep4ge targets TensorFlow/Keras FNN, CNN, and RNN programs. Transformers fall outside its scope. Stack Overflow programs may not reflect industrial development practices, and mutation-derived faults may not represent naturally occurring faults. The results therefore do not establish generalization to industrial programs or real-fault distributions. Correct baselines were validated for execution and training, not semantic correctness. Hardware-related features depend on the execution environment and may vary across platforms.

\section{Conclusion}
Deep4ge provides a controlled, mutation-based benchmark of 14,227 DNN training runs from adapted Stack Overflow programs. In our proof-of-concept baselines, per-epoch trajectory summaries produced higher MCC than final-epoch features for both fault detection and diagnosis. The dataset is publicly archived to support replication and future extension within its documented scope.

\bibliographystyle{IEEEtran}
\bibliography{references}

\begin{thebibliography}{10}
\providecommand{\url}[1]{#1}
\csname url@samestyle\endcsname
\providecommand{\newblock}{\relax}
\providecommand{\bibinfo}[2]{#2}
\providecommand{\BIBentrySTDinterwordspacing}{\spaceskip=0pt\relax}
\providecommand{\BIBentryALTinterwordstretchfactor}{4}
\providecommand{\BIBentryALTinterwordspacing}{\spaceskip=\fontdimen2\font plus
\BIBentryALTinterwordstretchfactor\fontdimen3\font minus \fontdimen4\font\relax}
\providecommand{\BIBforeignlanguage}[2]{{%
\expandafter\ifx\csname l@#1\endcsname\relax
\typeout{** WARNING: IEEEtran.bst: No hyphenation pattern has been}%
\typeout{** loaded for the language `#1'. Using the pattern for}%
\typeout{** the default language instead.}%
\else
\language=\csname l@#1\endcsname
\fi
#2}}
\providecommand{\BIBdecl}{\relax}
\BIBdecl

\bibitem{Zhang2021AutoTrainer}
X.~Zhang, J.~Zhai, S.~Ma, and C.~Shen, ``{AutoTrainer}: An automatic {DNN} training problem detection and repair system,'' in \emph{ICSE}.\hskip 1em plus 0.5em minus 0.4em\relax IEEE, 2021, pp. 359--371.

\bibitem{Schoop2021Umlaut}
E.~Schoop, F.~Huang, and B.~Hartmann, ``{UMLAUT}: Debugging deep learning programs using program structure and model behavior,'' in \emph{CHI}.\hskip 1em plus 0.5em minus 0.4em\relax ACM, 2021, pp. 1--16.

\bibitem{Wardat2021DeepLocalize}
M.~Wardat, W.~Le, and H.~Rajan, ``{DeepLocalize}: Fault localization for deep neural networks,'' in \emph{ICSE}.\hskip 1em plus 0.5em minus 0.4em\relax IEEE, 2021, pp. 251--262.

\bibitem{Wardat2022DeepDiagnosis}
M.~Wardat, B.~D. Cruz, W.~Le, and H.~Rajan, ``{DeepDiagnosis}: Automatically diagnosing faults and recommending actionable fixes in deep learning programs,'' in \emph{ICSE}.\hskip 1em plus 0.5em minus 0.4em\relax IEEE, 2022, pp. 561--572.

\bibitem{Cao2022DeepFD}
J.~Cao, M.~Li, X.~Chen, M.~Wen, Y.~Tian, B.~Wu, and S.~Cheung, ``{DeepFD}: Automated fault diagnosis and localization for deep learning programs,'' in \emph{ICSE}.\hskip 1em plus 0.5em minus 0.4em\relax IEEE, 2022, pp. 573--585.

\bibitem{Humbatova2021DeepCrime}
N.~Humbatova, G.~Jahangirova, and P.~Tonella, ``{DeepCrime}: Mutation testing of deep learning systems based on real faults,'' in \emph{ISSTA}.\hskip 1em plus 0.5em minus 0.4em\relax ACM, 2021, pp. 67--78.

\bibitem{islam_comprehensive_2019}
M.~J. Islam, G.~Nguyen, R.~Pan, and H.~Rajan, ``A comprehensive study on deep learning bug characteristics,'' in \emph{ESEC/FSE}.\hskip 1em plus 0.5em minus 0.4em\relax ACM, 2019, pp. 510--520.

\bibitem{Jahangirova2020Mutation}
G.~Jahangirova and P.~Tonella, ``An empirical evaluation of mutation operators for deep learning systems,'' in \emph{ICST}.\hskip 1em plus 0.5em minus 0.4em\relax IEEE, 2020, pp. 74--84.

\bibitem{Jahan2025DEFault}
S.~Jahan, M.~B. Shah, P.~Mahbub, and M.~M. Rahman, ``Improved detection and diagnosis of faults in deep neural networks using hierarchical and explainable classification,'' in \emph{ICSE}.\hskip 1em plus 0.5em minus 0.4em\relax IEEE, 2025, pp. 2944--2956.

\bibitem{Breiman2001RandomForests}
L.~Breiman, ``Random forests,'' \emph{Machine Learning}, vol.~45, no.~1, pp. 5--32, 2001.

\bibitem{Matthews1975MCC}
B.~W. Matthews, ``Comparison of the predicted and observed secondary structure of {T4} phage lysozyme,'' \emph{Biochimica et Biophysica Acta (BBA) -- Protein Structure}, vol. 405, no.~2, pp. 442--451, 1975.

\bibitem{Chicco2020MCC}
D.~Chicco and G.~Jurman, ``The advantages of the {Matthews} correlation coefficient (mcc) over {F1} score and accuracy in binary classification evaluation,'' \emph{BMC Genomics}, vol.~21, no.~1, p.~6, 2020.

\bibitem{grossberg2013recurrent}
S.~Grossberg, ``Recurrent neural networks,'' \emph{Scholarpedia}, vol.~8, no.~2, p. 1888, 2013.

\bibitem{Ma2018DeepMutation}
L.~Ma, F.~Zhang, J.~Sun, M.~Xue, B.~Li, F.~Juefei{-}Xu, C.~Xie, L.~Li, Y.~Liu, J.~Zhao, and Y.~Wang, ``{DeepMutation}: Mutation testing of deep learning systems,'' in \emph{ISSRE}.\hskip 1em plus 0.5em minus 0.4em\relax IEEE, 2018, pp. 100--111.

\bibitem{Hu2019DeepMutationPP}
Q.~Hu, L.~Ma, X.~Xie, B.~Yu, Y.~Liu, and J.~Zhao, ``{DeepMutation++}: A mutation testing framework for deep learning systems,'' in \emph{ASE}.\hskip 1em plus 0.5em minus 0.4em\relax IEEE, 2019, pp. 1158--1161.

\end{thebibliography}

\end{document}